# Spectroscopy of f-f transitions, crystal-field calculations, and magnetic and quadrupole helix chirality in DyFe$_3$(BO$_3$)$_4$

M. N. Popova, E. P. Chukalina, and K. N. Boldyrev*

*Institute of Spectroscopy, Russian Academy of Sciences, 108840 Moscow, Troitsk, Russia*

T. N. Stanislavchuk

*Department of Physics, New Jersey Institute of Technology, Newark, NJ 07102 USA*

B. Z. Malkin

*Kazan Federal University, 420008 Kazan, Russia*

I. A.Gudim

*Kirensky Institute of Physics, Siberian Branch of RAS, 660036 Krasnoyarsk, Russia*

Recently, quadrupole helix chirality and its domain structure was observed in resonant x-ray diffraction experiments on DyFe$_3$(BO$_3$)$_4$ using circularly polarized x rays [T. Usui et al., Nature Materials **13**, 611 (2014)]. We show that this effect can be explained quantitatively by calculating the quadrupole moments of the Dy$^{3+}$ ions induced by the low-symmetry ($C_2$) crystal-field (CF) component. In this work, the CF parameters for the Dy$^{3+}$ ions in the $P3_121$ ($P3_221$) phase of DyFe$_3$(BO$_3$)$_4$ are obtained from CF calculations based on the analysis of high-resolution temperature-dependent optical spectroscopy data. We also consider the helix chirality of the single-site magnetic susceptibility tensors of the Dy$^{3+}$ ions in the paramagnetic $P3_121$ ($P3_221$) phase and suggest a neutron diffraction experiment to reveal it.

*Corresponding author*. E-mail address: *kn.boldyrev@gmail.com*

## I. INTRODUCTION

Chirality or handedness is the asymmetry of an object upon its mirroring. In crystals, chirality can be connected with a helical crystal structure, which gives rise to such functionality as optical rotatory power [1]. Magnetic chirality corresponding to twisted magnetic structures induces ferroelectricity in a number of multiferroics [2]. Recent work [3] has introduced a new concept of "chirality of electronic quadrupole moments", on the base of resonant x-ray diffraction studies using circularly polarized x rays, which may lead to a new material functionality. In Ref. [3], single-crystal specimens of DyFe$_3$(BO$_3$)$_4$ were employed.

The DyFe$_3$(BO$_3$)$_4$ crystals belong to the borate family with general formula $RM_3$(BO$_3$)$_4$ where $R$ is a rare earth or Y ion but $M$ = Al, Ga, Sc, Fe, or Cr. These compounds crystallize in the trigonal $R32$ structure of the natural mineral huntite [4]. The most known representatives of this family are aluminum borates, which are used in self-frequency doubling and self-frequency summing lasers (see, e.g., Refs [5-7] and references therein). Additional interest of scientists in the huntite borates is connected with an appreciable magnetoelectric effect found in the yttrium and rare-earth (RE) iron borates (see, e.g., Refs [8-11] and references therein) and in the RE aluminum and gallium borates [12-15].

With decreasing the temperature, the iron borates with $R$=Eu-Er or Y undergo a structural phase transition from the $R32$ ($D_3^7$, №155) phase to a less symmetric but also trigonal one [16-18] corresponding to the enantiomorphic space-group pair $P3_121$ ($D_3^4$, №152) and $P3_221$ ($D_3^6$, №154). Structural motives showing a difference between these two groups are presented in Fig. 1. Resonant x-ray diffraction experiments on single-crystal samples of DyFe$_3$(BO$_3$)$_4$ [3] revealed macroscopic domains with (sub)millimeter dimensions, which differ by right-handed ($P3_121$) or left-handed ($P3_221$) helical structures of electronic quadrupole moments of the Dy$^{3+}$ ions at temperatures below the structural transition temperature $T_S$=285 K [3]. The domain pattern is robust against passing $T_S$ several times. This is because the high-temperature crystallographic structure ($R32$) is already chiral, so that both left- and right-handed crystallographic domains containing left- and right-handed helix chains of FeO$_6$ octahedra, respectively, already exist above $T_S$. The quadrupole helix chirality of Dy$^{3+}$ ions develops below $T_S$

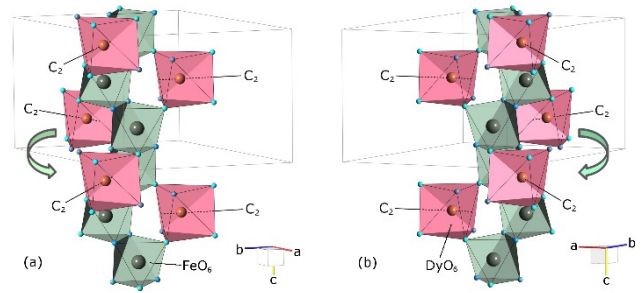

FIG. 1. (a) Left-handed $P3_221$ and (b) right-handed $P3_121$ helical structures of DyO$_6$ distorted prisms and FeO$_6$ octahedra in DyFe$_3$(BO$_3$)$_4$ crystal at temperatures below the structural transition temperature $T_S$=285 K. Three crystallographically equivalent but magnetically non-equivalent Dy ions with the local $C_2$ symmetry are connected by the (a) left-handed and (b) right-handed ($2\pi/3$) rotations around the $c$-axis.

(and coexists with the antiferromagnetic order that sets below $T_N$=39 K [19,20]), being closely connected with the crystallographic chirality above $T_S$. The authors of Ref. [3] conjectured that, because of a significant spin-orbit interaction, the direction of a magnetic moment at the Dy site in DyFe$_3$(BO$_3$)$_4$ is strongly coupled with the orientation of its electric quadrupole. It is important to note that a multidomain chiral structure may noticeably reduce multiferroic properties of a crystal. It has been found, in particular, that the value of the spontaneous electric polarization observed in SmFe$_3$(BO$_3$)$_4$ below $T_N$ appreciably depends on a sample [21]. Control of quadrupole helix chirality using circularly polarized x-rays might help to develop growth technologies for obtaining single-domain crystals.

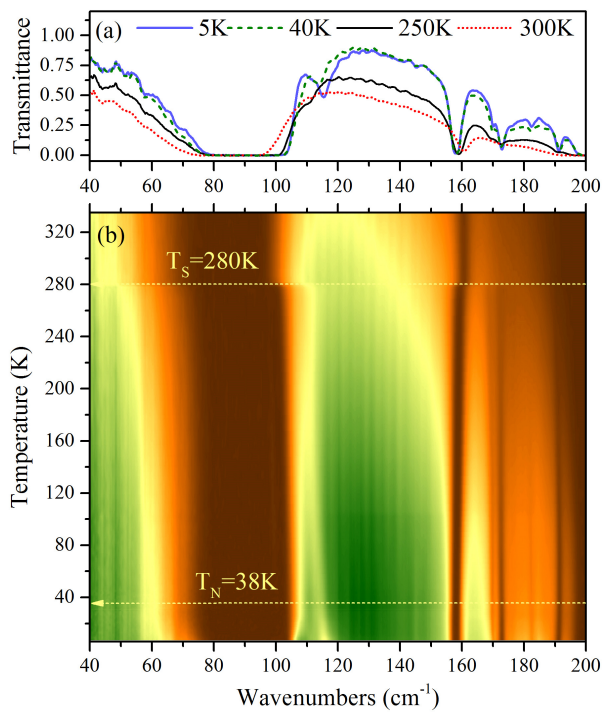

FIG. 2. Far-infrared α-polarized transmission spectra of a DyFe$_3$(BO$_3$)$_4$ single crystal (a) at several selected temperatures; (b) presented as intensity maps in the wavenumber – temperature scales.

In the present work, we show that electric quadrupole moments of Dy$^{3+}$ ions in DyFe$_3$(BO$_3$)$_4$ are connected with the low-symmetry crystal-field (CF) component that appears below the temperature $T_S$ of the structural phase transition, at which the point symmetry group of the Dy site lowers from $D_3$ to $C_2$. A limited available information on CF levels of Dy$^{3+}$ ions in DyFe$_3$(BO$_3$)$_4$ [22] prevents from doing crystal-field calculations in the frame of $C_2$ symmetry CF Hamiltonian with 15 parameters. In Ref. [23], CF parameters for DyFe$_3$(BO$_3$)$_4$ were estimated from the fit to the experimental magnetization data assuming the $D_3$ symmetry. However, the values and signs of some so obtained CF parameters contradict to general trends in variations of the CF parameters, even in the RE borates with the $R32$ structure [24-26]. To find reliable CF parameters, we perform a thorough broadband high-resolution optical spectroscopy investigation of the dysprosium iron borate. On the basis of a large set of experimentally found CF energies and preliminary calculations in the framework of the exchange-charge model [27], we obtain CF parameters and wave functions of the Dy$^{3+}$ ions in the low-symmetry ($P3_121$ or $P3_221$) phase of DyFe$_3$(BO$_3$)$_4$, calculate the electric quadrupole moments, and compare them with those found experimentally in Ref. [3]. We also calculate values of the single-site magnetic susceptibility tensors of the Dy$^{3+}$ ions in the paramagnetic $P3_121$ ($P3_221$) phase, consider the helix chirality of these tensors, and suggest an experiment that could bring to light a magnetic chirality in the low-symmetry paramagnetic phase of DyFe$_3$(BO$_3$)$_4$.

## II. EXPERIMENTAL

Single crystals of dysprosium iron borate were grown on the seeds from solution-melts on the base of Bi$_2$Mo$_3$O$_{12}$, as described in Refs. [28,29]. To prepare samples for spectroscopic measurements, we cut out 0.2-1.3 mm thick plates either perpendicular or parallel to the crystallographic $c$ axis of the crystal. Transmission spectra were detected in wide spectral (3000-23000 cm$^{-1}$) and temperature (3.5-300 K) regions using Bruker 125HR and Bomem DA3.002 Fourier spectrometers, with a spectral resolution up to 0.5 cm$^{-1}$. The spectra were registered in different polarizations of the incident radiation, namely, in σ ($\mathbf{k}\perp c$, $\mathbf{E}\perp c$), π ($\mathbf{k}\perp c$, $\mathbf{E}\parallel c$), and α ($\mathbf{k}\parallel c$, $\mathbf{E}$, $\mathbf{H}\perp c$) polarizations. A very weak optical transition to the $^6F_{1/2}$ level of Dy$^{3+}$ was detected using a big nonoriented sample and nonpolarized light. For cryogenic measurements, we used either an optical helium-vapor cryostat or a closed-cycle Cryomech ST-403 cryostat.

To characterize our samples, we studied their temperature-dependent (10 – 400 K) polarized infrared vibrational spectra. Being compared with the spectra of GdFe$_3$(BO$_3$)$_4$ [30], the crystal structure of which is well known from the temperature-dependent x-ray diffraction studies [16], these spectra unambiguously testify the $R32$ structure at high temperatures, the phase transition into the $P3_121$ ($P3_221$) structure at $T_S$=280 K, and an absence of any other phases in the samples. Figure 2 shows a part of the far-infrared transmission spectrum of DyFe$_3$(BO$_3$)$_4$. An abrupt shift of phonon lines and appearance of new phonons at $T_S$=280 K are due to a weak first-order structural phase transition [17,30].

In the high-temperature $R32$ structure, the crystal field of the $D_3$ symmetry splits multiplets of the Dy$^{3+}$ free ion with the $4f^9$ ground electronic configuration into $(2J+1)/2$ sublevels, where $J$ is the corresponding total half-integer angular moment. Each of these sublevels is doubly degenerated, this Kramers degeneracy can be removed by a magnetic field only. Thus, at the structural phase transition $R32 \rightarrow P3_121$ or $P3_221$, when the symmetry of the Dy$^{3+}$ ion position lowers from $D_3$ to $C_2$, all CF sublevels of the $Dy^{3+}$ ion remain doubly degenerated. A magnetic ordering at $T_N$=38 K was detected from the observed splitting of Kramers doublets in the temperature-dependent electronic spectra of $f$-$f$ transitions of $Dy^{3+}$ in $DyFe_3(BO_3)_4$ (not shown).

## III. EXPERIMENTAL RESULTS

In the transmission spectra of rare-earth iron borates, relatively narrow spectral lines corresponding to $f$-$f$ transitions in the RE ions are superimposed onto broad $d$-$d$ bands of the $Fe^{3+}$ ions [24,26]. After subtracting the $d$-$d$ absorption bands (e.g., measured in $GdFe_3(BO_3)_4$ where $f$-$f$ transitions start at energies above ~ 30000 cm$^{-1}$), one gets the spectra of the $f$-$f$ transitions. Figure 3 shows thus obtained absorption spectrum of $DyFe_3(BO_3)_4$ at $T$=50 K in the spectral range 2000-15000 cm$^{-1}$. As one can see in Figure 2, the first excited multiplet $^6H_{13/2}$ is separated from the ground multiplet $^6H_{15/2}$ by an energy gap of ~3500 cm$^{-1}$ (~5000 K). Thus, only the CF levels of the ground multiplet are populated in the studied temperature range (3.5-300 K).

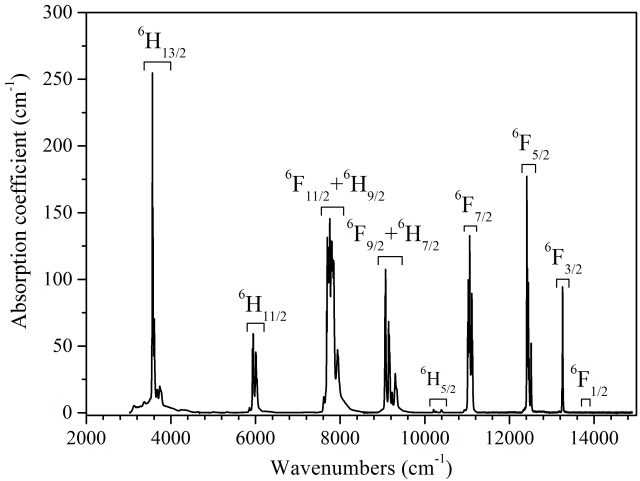

FIG. 3. The σ-polarized absorption spectrum of the $Dy^{3+}$ ions in the dysprosium iron borate at $T$=50 K.

Since physical properties of a system are determined by its populated states, data on CF levels of the ground multiplet are of main interest. Figure 4 presents at expanded scales absorption spectra in the regions of transitions from the ground multiplet $^6H_{15/2}$ to different excited multiplets of the $Dy^{3+}$ ion. Here and hereafter, notation of a spectral line contain the initial and final levels of an optical transition; CF levels of the ground multiplet are labeled by numerals and those of a given excited multiplet – by capital letters (A, B, etc.). The lines designated 2A, 2B, etc. diminish in intensity with further lowering the temperature until they completely disappear, while the lines 1A, 1B, etc., which are shifted by 15 cm$^{-1}$ to higher frequencies from the lines 2A, 2B, etc., grow in intensity. Such behavior shows unambiguously that the spectral lines observed at 50 K correspond to the transitions from the lowest two sublevels of the ground multiplet $^6H_{15/2}$ separated by 15 cm$^{-1}$ and populated at this temperature. With increasing the temperature, spectral lines due to transitions from higher excited levels of the $^6H_{15/2}$ multiplet appear in the spectra. These lines are rather weak, and to improve signal to noise ratio, spectra at high temperatures were registered on the samples of larger thickness. The obtained energies of the excited CF sublevels of the ground multiplet $^6H_{15/2}$ are presented in Table I.

The optical multiplets measured in $DyFe_3(BO_3)_4$ with the low-temperature structure ($P3_121$ or $P3_221$) are similar to those measured in $YAl_3(BO_3)_4$ crystals (the $R32$ space group) containing impurity $Dy^{3+}$ ions at $Y^{3+}$ sites with the local $D_3$ symmetry [31-33]. However, the total splittings of dysprosium borate having larger lattice constants ($a$=0.95439 and 0.92833 nm, $c$=0.75676 and 0.72331 nm at room temperature in DyFe$_3$(BO$_3$)$_4$ [20] and YAl$_3$(BO$_3$)$_4$ [34,35], respectively). Almost all the spectral lines observed in the transmission and absorption spectra of the paramagnetic phase of DyFe$_3$(BO$_3$)$_4$, in particular, at $T$=50 K, were unambiguously identified, and the corresponding energies of excitations of the Dy$^{3+}$ ions in the range 3500-22400 cm$^{-1}$ are given in Table I (columns 2 and 5). This dataset is a basis for the CF modeling described in the next section. Shifts and splittings of the spectral lines observed at temperatures below the magnetic ordering temperature $T_\text{N}$=39 K will be discussed in a separate publication.

TABLE I. Crystal-field energies $E$ (cm$^{-1}$) of the $Dy^{3+}$ ion in the paramagnetic ($T$=50 K) phase of $DyFe_3(BO_3)_4$ determined from the analysis of the spectra and from calculations.

| $^{2S+1}L_J$ | | $E$ | | $^{2S+1}L_J$ | | $E$ | |
|---|---|---|---|---|---|---|---|
| | | Experiment | Theory | | | Experiment | Theory |
| 1 | | 2 | 3 | 4 | | 5 | 6 |
| $^6H_{15/2}$ | 1 | 0 | 0 | | A | 9170 | 9161.4 |
| | 2 | 15 | 15.6 | | B | 9228 | 9231.2 |
| | 3 | 97 | 94.5 | $^6F_{9/2}$ | C | 9301 | 9298.8 |
| | 4 | 186 | 191.4 | | D | - | 9308.2 |
| | 5 | 228 | 231.5 | | E | 9337 | 9333.2 |
| | 6 | 270 | 280.3 | | | | |
| | 7 | 335 | 345 | | A | 10219 | 10227 |
| | 8 | - | 377.4 | $^6H_{5/2}$ | B | 10263.5 | 10273 |
| | | | | | C | 10406 | 10400 |
| $^6H_{13/2}$ | A | 3567 | 3571.4 | | | | |
| | B | 3579.6 | 3580.5 | | A | 11027.2 | 11033 |
| | C | 3610 | 3608.5 | $^6F_{7/2}$ | B | 11059.2 | 11057 |
| | D | 3626.2 | 3633 | | C | 11089.2 | 11071 |
| | E | 3688 | 3684.2 | | D | 11124.8 | 11098 |
| | F | 3741 | 3732.6 | | | | |
| | G | 3782.7 | 3773.3 | | A | 12422.3 | 12425 |
| | | | | $^6F_{5/2}$ | B | 12459.5 | 12455 |
| | A | 5858 | 5860 | | C | 12509.6 | 12491 |
| | B | 5947 | 5951.6 | | | | |
| $^6H_{11/2}$ | C | 5977.3 | 5976.8 | $^6F_{3/2}$ | A | 13252.4 | 13246 |
| | D | 6004.4 | 6008.4 | | B | 13262.0 | 13256 |
| | E | 6013 | 6015.7 | | | | |
| | F | 6037.8 | 6044.1 | $^6F_{1/2}$ | A | 13806.5 | 13800 |
| | A | 7696 | 7703.5 | | A | 21008.6 | 21011 |
| | B | 7711.3 | 7716.5 | | B | 21077.3 | 21081 |
| $^6H_{9/2}$ | C | 7730 | 7732.9 | $^4F_{9/2}$ | C | - | 21100 |
| | D | 7756 | 7762.3 | | D | 21255.0 | 21233 |
| | E | 7799 | 7807 | | E | - | 21244 |
| | A | 7820 | 7819.1 | | A | 22075 | 22074 |
| | B | 7847 | 7839.6 | | B | 22093 | 22092 |
| $^6F_{11/2}$ | C | 7867 | 7877.7 | | C | 22127 | 22139 |
| | D | 7921 | 7932.9 | $^4I_{15/2}$ | D | - | 22203 |
| | E | 7954 | 7954 | | E | 22254 | 22243 |
| | F | - | 7976.2 | | F | 22298 | 22284 |
| | | | | | G | 22348 | 22339 |
| | A | 9062 | 9058.3 | | H | - | 22375 |
| $^6H_{7/2}$ | B | 9073 | 9064.7 | | | | |
| | C | 9102 | 9101.9 | | | | |
| | D | 9158 | 9150.3 | | | | |

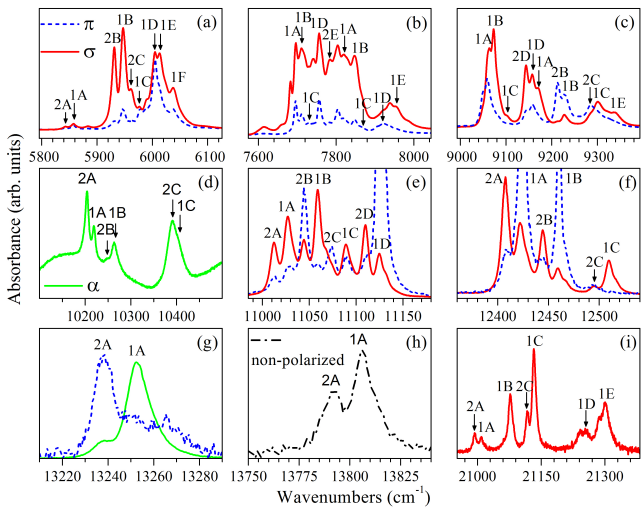

FIG. 4. Absorption spectra of DyFe$_3$(BO$_3$)$_4$ at the temperature 50 K in the σ (solid red lines), π (dashed blue lines) and α (solid green lines) polarizations as well as in nonpolarized radiation (black dash-dot lines), corresponding to transitions from the ground multiplet $^6$H$_{15/2}$ to sublevels of the excited multiplets (a) $^6$H$_{11/2}$, (b) $^6$H$_{9/2}$ and $^6$F$_{11/2}$, (c) $^6$H$_{7/2}$ and $^6$F$_{9/2}$, (d) $^6$H$_{5/2}$, (e) $^6$F$_{7/2}$, (f) $^6$F$_{5/2}$, (g) $^6$F$_{3/2}$, (h) $^6$F$_{1/2}$, and (i) $^4$F$_{9/2}$.

## IV. DISCUSSION

### A. Crystal-field calculations for the $P3_121$ ($P3_221$) paramagnetic phase of DyFe$_3$(BO$_3$)$_4$

The primitive cell of the RE iron borate crystal lattice in the $P3_121$ (or $P3_221$) phase contains three formula units. Correspondingly, there are three crystallographically equivalent but magnetically non-equivalent RE ions with the local $C_2$ symmetry, connected by the ($2\pi/3$) rotations around the $c$-axis (see Fig. 1) and two types of nonequivalent helicoidal chains of the FeO$_6$ octahedra containing the Fe$^{3+}$ ions at sites with the $C_2$ (Fe1) and $C_1$ (Fe2) local symmetries, respectively. The $C_2$ symmetry axes lie in the $ab$ plane. In this study, when analyzing spectral and static magnetic properties of DyFe$_3$(BO$_3$)$_4$, we assume that electrons are localized at the ions that form the crystal lattice and, following the approach used by us earlier in Refs. [24-26, 36] start from a consideration of independent Dy$^{3+}$ ions and dimers containing the nearest-neighbor iron ions (Fe$^{3+}$)$_2$ in the chains coupled by the isotropic exchange. At this stage, all interactions between the ions, which involve dynamic electronic variables, even relative the time inversion, are considered in the framework of the CF approximation. A single Dy$^{3+}$ ion and an iron dimer are described by the corresponding effective Hamiltonians. Further, the exchange interactions between the Fe$^{3+}$ ions and between the Fe$^{3+}$ and Dy$^{3+}$ ions are taken into account in the framework of the self-consistent-field approximation.

The structure of the spectrum of an isolated Dy$^{3+}$ ion with the ground 4$f^9$ electronic configuration in a dielectric crystal can be described using the Hamiltonian

$$H_0 = H_\text{FI} + H_\text{CF}, \quad (1)$$

where

$$H_\text{FI} = \zeta \sum_j \mathbf{l}_j \mathbf{s}_j + \alpha \hat{L}^2 + \beta \hat{G}(G_2) + \gamma \hat{G}(R_7) + \sum_q (F^q \hat{f}_q + P^q \hat{p}_q + T^q \hat{t}_q + M^q \hat{m}_q) \quad (2)$$

is the free-ion Hamiltonian written in a standard form that includes the energies of the spin-orbit and electrostatic interactions between the 4$f$ electrons and additional terms due to interconfigurational interactions [37]. The Hamiltonian $H_\text{CF}$ stands for the energy of localized 4$f$ electrons (labeled by the index $j$) in a static crystal field. Here, $\mathbf{l}_j$ and $\mathbf{s}_j$ are orbital and spin moments of electrons, $\hat{L}$ is the total orbital moment, the explicit forms of operators $\hat{G}$, $\hat{f}$, $\hat{p}$, $\hat{t}$, $\hat{m}$ are determined in literature (see references in [37]). In the present work, we use parameters of the Hamiltonian (2) from Ref. [35] for the impurity Dy$^{3+}$ ions in YAl$_3$(BO$_3$)$_4$, slightly corrected to fit gaps between multiplet centers of gravity: $F^2$=91060, $F^4$=63871, $F^6$=49460, $\zeta$=1909, $\alpha$=18, $\beta$=-633, $\gamma$=1790, $P^2$=719, $P^4$=360, $P^6$=72, $M^0$=3.39, $M^2$=1.9, $M^4$=1.05, $T^2$=329, $T^3$=36, $T^4$=127, $T^6$=-314, $T^7$=404, $T^8$= 315 (in cm$^{-1}$).

TABLE II. Distances (in Angstroms) between the RE$^{3+}$ ion and the neighbor oxygen ions in the $P3_121$ ($P3_221$) phase of rare-earth iron borates.

| Rare-earth | $T$ | $R$(RE-O3) | $R$(RE-O4) | $R$(RE-O7) | Reference |
|---|---|---|---|---|---|
| Gd(4$f^7$) | 90 K | 2.3485 | 2.3830 | 2.3474 | [16] |
| Tb(4$f^8$) | 40 K | 2.3147 | 2.3893 | 2.3598 | [41] |
| Dy(4$f^9$) | 50 K | 2.3725 | 2.3520 | 2.3474 | [20] |
| Dy(4$f^9$) | 60 K | 2.3353 | 2.3838 | 2.3280 | [3] |
| Ho(4$f^{10}$) | 50 K | 2.3073 | 2.3938 | 2.3292 | [42] |

Let us introduce a local right-handed Cartesian system of coordinates with the origin at a selected Dy$^{3+}$ ion and with the $z_l$ axis along the crystallographic $c$ axis and the $x_l$ axis along the $C_2$ symmetry axis for the given site. In this local system of coordinates, the CF Hamiltonian for the Dy$^{3+}$ ion is determined by fifteen real CF parameters $B_q^p$:

$$H_{CF} = \sum_{j=1}^{9}\{B_0^2 C_0^{(2)}(j) + B_0^4 C_0^{(4)}(j) + B_0^6 C_0^{(6)}(j) + iB_{-3}^4[C_{-3}^{(4)}(j) + C_3^{(4)}(j)]$$

$$+ iB_{-3}^6[C_{-3}^{(6)}(j) + C_3^{(6)}(j)] + B_6^6[C_{-6}^{(6)}(j) + C_6^{(6)}(j)]$$

$$+ iB_{-1}^2[C_{-1}^{(2)}(j) + C_1^{(2)}(j)] + B_2^2[C_{-2}^{(2)}(j) + C_2^{(2)}(j)]$$

$$+ iB_{-1}^4[C_{-1}^{(4)}(j) + C_1^{(4)}(j)] + B_2^4[C_{-2}^{(4)}(j) + C_2^{(4)}(j)] \quad (3)$$

$$+ iB_{-1}^6[C_{-1}^{(6)}(j) + C_1^{(6)}(j)] + B_2^6[C_{-2}^{(6)}(j) + C_2^{(6)}(j)]$$

$$+ B_4^4[C_{-4}^{(4)}(j) + C_4^{(4)}(j)] + B_4^6[C_{-4}^{(6)}(j) + C_4^{(6)}(j)]$$

$$+ iB_{-5}^6[C_{-5}^{(6)}(j) + C_5^{(6)}(j)]\},$$

where $C_q^{(p)}$ is the spherical tensor operator of the rank $p$, the summation is over 4f electrons localized at the $Dy^{3+}$ ion. Six parameters, namely, $B_0^2$, $B_0^4$, $B_{-3}^4$, $B_0^6$, $B_{-3}^6$, and $B_6^6$ in the first two lines of Eq. (3), define the dominant CF component of trigonal symmetry and other nine parameters define the CF component of the $C_2$ symmetry. The initial values of the CF parameters were calculated in the framework of the exchange-charge model (ECM) [27]:

$$B_q^p = B_q^{(pc)p} + B_q^{(ec)p}. \quad (4)$$

The electrostatic field of point charges $eq_L$ ($e$ is the elemental charge) of lattice ions $L$ with the spherical coordinates $R_L$, $\theta_L$, $\varphi_L$ (the origin of the coordinates is at the $Dy^{3+}$ ion) is described by the parameters

$$B_q^{(pc)p} = -\sum_L e^2 q_L (1-\sigma_p) <4f|r^p|4f>(-1)^q C_{-q}^{(p)}(\vartheta_L,\phi_L) / R_L^{p+1}, \quad (5)$$

where $\sigma_p$ are the shielding factors [38], and $-e<4f|r^p|4f>$ are the moments of the 4f electron charge distribution. The interaction of 4f electrons with the field of exchange charges defined through the overlap integrals $S_s(R_L) = <4f0|2s>$, $S_\sigma(R_L) = <4f0|2p0>$, $S_\pi(R_L) = <4f1|2p1>$ between the 4f wave functions ($|4f \, l_z>$) of the $Dy^{3+}$ ion and 2s, 2p wave functions of the electrons at the outer filled electronic shells of the nearest oxygen ions is described by the parameters

$$B_q^{(ec)p} = \sum_L \frac{2(2p+1)}{7} \frac{e^2}{R_L} S_p(R_L)(-1)^q C_{-q}^{(p)}(\vartheta_L,\phi_L). \quad (6)$$

In a general case, the model operates with three phenomenological parameters $G_s$, $G_\sigma$, $G_\pi$ in the linear combinations of the squared overlap integrals

$$S_p(R_L) = G_s[S_s(R_L)]^2 + G_\sigma[S_\sigma(R_L)]^2 + [2-p(p+1)/12]G_\pi[S_\pi(R_L)]^2. \quad (7)$$

The dependencies of the overlap integrals on the distance $R$ (in Angstroms) between the ions, computed with the available radial wave functions of the $Dy^{3+}$ [39] and $O^{2-}$ [40] ions, can be approximated by the functions $S_u = a_u \exp(-b_u R^{c_u})$ ($u=s$, $\sigma$ and $\pi$), where $a_s$=0.26533, $b_s$=0.859, $c_s$=1.5476; $a_\sigma$=0.07039, $b_\sigma$=0.2495, $c_\sigma$=2.2061; $a_\pi$=1.40205, $b_\pi$=2.2761, $c_\pi$=0.9358.

TABLE III. Crystal-field parameters $B_q^p$ (cm$^{-1}$) for RE iron borates with the $P3_121$ structure. In the $P3_221$ structure, the parameters with $q$=-1, -3 and -5 have opposite signs.

| p | q | EuFe$_3$(BO$_3$)$_4$ [43] | TbFe$_3$(BO$_3$)$_4$ [36] | DyFe$_3$(BO$_3$)$_4$ | [Present work] |
|---|---|---|---|---|---|
| 1 | 2 | 3 | 4* | 5 |
| 2 | 0 | 484 | 434 | 200 | 404 |
| 4 | 0 | -1255 | -1256 | -1255 | -1192 |
| 4 | -3 | 619 | 608 | 539 | 554.4 |
| 6 | 0 | 404 | 352 | 337 | 328 |
| 6 | -3 | 80 | 73 | 69 | 70.3 |
| 6 | 6 | 290 | 270 | 203 | 232 |
| 2 | -1 | 39 | 38 | 66 | 58.4 |
| 4 | -1 | -76 | -66 | -43 | -49.2 |
| 6 | -1 | -32 | -27 | -0.3 | -7.4 |
| 2 | 2 | 54 | 54 | 123 | 69.4 |
| 4 | 2 | 102 | 82 | 104 | 101.2 |
| 6 | 2 | -11 | -8 | -13 | -14 |
| 4 | 4 | -26 | -23 | 16 | 15.9 |
| 6 | 4 | -31 | -27 | 31 | 31.4 |
| 6 | -5 | -131 | -91 | -92 | -79 |

In rare-earth iron borates, interactions of RE ions with the nearest-neighbor oxygen ions bring the main contributions into the CF parameters. The lattice structure data for DyFe$_3$(BO$_3$)$_4$ were presented in Refs. [3] and [20]. In the $P3_121$ ($P3_221$) phase, the RE ions are at the 3a Wyckoff positions. Below, we use the same atomic notations as in Refs. [3, 20] and consider explicitly the reference $Dy^{3+}$ ion with the fractional coordinates (-$x_{Dy}$, -$x_{Dy}$, 0). The nearest- neighbor six oxygen ions at the 6c Wyckoff sites O3, O4, and O7 form a deformed trigonal prism and can be divided into three pairs with fractional coordinates ($x_{Oi}$, $y_{Oi}$, $z_{Oi}$) and ($y_{Oi}$, $x_{Oi}$, -$z_{Oi}$) at the distances $R$(RE-O$i$) from the RE ion ($i$=3, 4, 7). As is seen in Table II, the distances between the $Dy^{3+}$ and oxygen O3 and O4 ions presented in Ref. [20] contradict to trends in variations of these distances with the number of electrons in the 4f shell of a RE ion in other RE iron borates with the same structure. In the calculations described below, we use the structural data from Ref. [3] measured at the temperature 60 K, which seem to be more reliable. The transformation of x and y coordinates in the crystallographic trigonal frame presented in Ref. [3] to the Cartesian system of coordinates at the reference dysprosium site is defined by the equations

$$x_l = (x - y/2)\cos(\pi/3) + (\sqrt{3}y/2)\sin(\pi/3), \quad (8)$$

$$y_l = -(x - y/2)\sin(\pi/3) + (\sqrt{3}y/2)\cos(\pi/3). \quad (9)$$

The results of calculations of the CF parameters (4) performed in the framework of the simplest single-parameter version of ECM ($G_s$=$G_\sigma$=$G_\pi$=7.5) are presented in Table III (column 4). The following values were used in calculations: $<r^2>$= 0.726, $<r^4>$=1.322, and $<r^6>$=5.107 (in atomic units) [39]; $\sigma_2$=0.646 [38], $\sigma_4$=$\sigma_6$=0; $q_{Dy}$=3, $q_{Fe}$=2, $q_B$=2.25, $q_O$=-1.5 (the values of the ion charges). The obtained CF parameters correlate well with the results of similar calculations of the CF parameters for the $Tb^{3+}$ ions in terbium iron borate [36] carried out with the ECM parameters $G_s = G_\sigma = G_\pi = 7$, using the structure data from Ref. [41] for 40 K.

Next, the energies of transitions between the energy levels of the $Dy^{3+}$ ions, obtained from the numerical diagonalization of the Hamiltonian (1) operating in the total space of 2002 states of the electronic $4f^9$ configuration, are compared with the measured optical spectra of $DyFe_3(BO_3)_4$ in the paramagnetic phase, and the initial CF parameters are varied to fit the experimental data. The final set of the CF parameters is presented in Table III (column 5), the corresponding CF energies (columns 3 and 6 in Table I) match well the results of measurements (columns 2 and 5 in Table I). Note that values of the six CF parameters $B_0^2$, $B_0^4$, $B_{-3}^4$, $B_0^6$, $B_{-3}^6$, and $B_6^6$ which determine the trigonal component of the crystal field change monotonically along the series of RE iron borates (in particular, CF parameters for the europium, terbium and dysprosium compounds are compared in Table III; to add the data on Pr, Nd, and Sm compounds, see Table III of Ref. [26]). The remaining nine parameters of the CF component of the $C_2$ symmetry are determined with lower accuracy, because of their relatively small influence on the multiplet splittings. However, just this CF component determines some specific properties of RE-iron borates in the $P3_121$ (or $P3_221$) phase, in particular, the anisotropy of the magnetic spectroscopic factors of Kramers doublets of the $Dy^{3+}$ ions in the $ab$-plane and the quadrupole helix chirality in $DyFe_3(BO_3)_4$ [3].

### B. Components of the g-tensor for the $Dy^{3+}$ ions, paramagnetic susceptibility of $DyFe_3(BO_3)_4$, and helix chirality of the local $Dy^{3+}$ susceptibility

First of all, we discuss below peculiarities of the g tensor in the case of the low-symmetry $P3_121$ (or $P3_221$) phase. The effective spin-Hamiltonian for a Kramers doublet in a magnetic field $\boldsymbol{B}$ can be written as

$$H_S = \mu_B(g_1 S_1 B_1 + g_2 S_2 B_2 + g_3 S_3 B_3), \quad (10)$$

where $\mu_B$ is the Bohr magneton, $B_\alpha$ and $S_\alpha$ are projections of the magnetic field and effective spin moment ($S=1/2$) on the principal axes of the tensor $G = g \cdot \tilde{g}$, and $g_\alpha$ are the square roots of the corresponding eigenvalues of this tensor. Components of the g-tensor are determined by matrix elements of the magnetic moment operator of an ion, $\boldsymbol{m} = -\mu_B \sum_j (\xi \boldsymbol{l}_j + 2\boldsymbol{s}_j)$ (here $\xi$ is the orbital reduction factor), in the basis of the wave functions $|\pm>$ of a Kramers doublet:

$$\begin{aligned} g_{\alpha x} &= 2\mathrm{Re} <+|m_\alpha|->/\mu_B, \\ g_{\alpha y} &= -2\mathrm{Im} <+|m_\alpha|->/\mu_B, \\ g_{\alpha z} &= 2<+|m_\alpha|+>/\mu_B. \end{aligned} \quad (11)$$

In a crystal field of the $C_2$ symmetry, one of the principal axes is directed along the symmetry axis (i.e., the $x_l$ axis in our case). From calculations carried out in the basis of the eigenfunctions of the Hamiltonian (1), using the orbital reduction factor $\xi = 0.95$ (this value was found from the fit of the high temperature magnetic susceptibilities, see below), we obtained the following principal values of the g-tensors: $g_1 = g_{xx} = 0.072$ and 0.096, $g_2 = 0.005$ and 0.012, $g_3 = 15.874$ and 14.369 for the ground and the first excited doublets of the $Dy^{3+}$ ions in $DyFe_3(BO_3)_4$, respectively. Note that the calculated largest g-factors ($g_3$) agree well with the measured g-factors along the principal $c$-axis $g_{cc}$=15.78 and 12.03 of the ground and the first excited doublets of the impurity $Dy^{3+}$ ions in $YAl_3(BO_3)_4$ with the $R32$ structure [44]. However, the principal axes corresponding to these largest $g_3$ factors in $DyFe_3(BO_3)_4$ are declined from the $c$-axis towards the $y_l$ axis by the angles of 23.5° and 2.7° for the ground and the first excited doublets, respectively.

As a consequence, contrary to the RE iron (and aluminum) borates in the $R32$ phase, in paramagnetic borates with the $P3_121$ (or $P3_221$) structure, the external magnetic field parallel to the $c$-axis induces not only longitudinal, but also local transversal components of the magnetic moments; similarly, the magnetic field oriented in the $ab$-plane induces local magnetic moments with projections on the $c$-axis of different signs and values at different RE sites, depending on the angle between the field direction and the local $C_2$ axis.

Further, we calculate static magnetic susceptibilities $\chi_\parallel$ and $\chi_\perp$ of $DyFe_3(BO_3)_4$ single crystals for external magnetic fields parallel and perpendicular to the $c$-axis, respectively, as functions of the temperature in a paramagnetic range $T_N < T < T_s$. In the paramagnetic phase, the renormalization of the susceptibility of the $Fe^{3+}$ ions due to the $f$-$d$ exchange interaction is rather weak (see Refs. [24,25,36]) and we compare the sums of the computed susceptibilities of the independent rare-earth and iron subsystems with the experimental data [19] in Fig. 5. It should be noted that we neglect here differences between the magnetic properties of the Fe1 and Fe2 ions. The principal axes of the susceptibility tensors of the three magnetically non-equivalent $Dy^{3+}$ ions in the unit cell rotate around the $c$-axis in accordance with the rotations of the local systems of coordinates introduced above. Correspondingly, all the ions have the same projections of their magnetic moments on the external field parallel to the $c$-axis, and $\chi_\parallel = N_A(\chi_{Dy,zz} + 3\chi_{Fe})$, where $N_A$ is the Avogadro number. The transversal component of the bulk susceptibility tensor is written as follows: $\chi_\perp = N_A[(\chi_{Dy,xx} + \chi_{Dy,yy})/2 + 3\chi_{Fe}]$. The effective isotropic susceptibility $\chi_{Fe}$ of an iron ion was calculated in the framework of the dimer model [24] using the values of the exchange integrals $J_{nn}$=-8 K and $J_{nnn}$=-2.5 K for the nearest-neighbor and next-nearest-neighbor antiferromagnetic interactions between the iron ions within the chains and in the neighboring chains, respectively. These values are slightly larger than the obtained earlier exchange integrals in terbium and samarium iron borates [26,36], due to shorter distances between the ions in dysprosium iron borate. The components of the susceptibility tensor for a single $Dy^{3+}$ ion were computed vs temperature with a step of one K, as quantum statistical averages of the magnetic moment **m** using the equilibrium density matrix defined by the Hamiltonian (1) added by the interaction of the $Dy^{3+}$ ion with a magnetic field $B$=0.1 T. A good agreement between the calculated and measured temperature dependencies of the longitudinal and transversal susceptibilities (see Fig.5) testifies a reliability of the parameters derived in the present work.

TABLE IV. Quadrupole moments of the $Dy^{3+}$ ions in the $P3_121$ phase of $DyFe_3(BO_3)_4$.

| Temperature | $<Q_{x_l^2-y_l^2}>$ | | $<Q_{y_l z_l}>/<Q_{x_l^2-y_l^2}>$ | |
|---|---|---|---|---|
| | Measured [3]* | Computed** | Measured [3] | Computed |
| 50 K | -1.22 | -1.18 | -1.23 | -1.415 |
| 100 K | -0.906 | -0.906 | -1.20 | -1.202 |
| 150 K | -0.678 | -0.695 | -1.10 | -1.096 |
| 200 K | -0.50 | -0.548 | -0.90 | -1.031 |

\* Arbitrary units
\*\*The absolute values are scaled to match the measured value at $T$=100 K

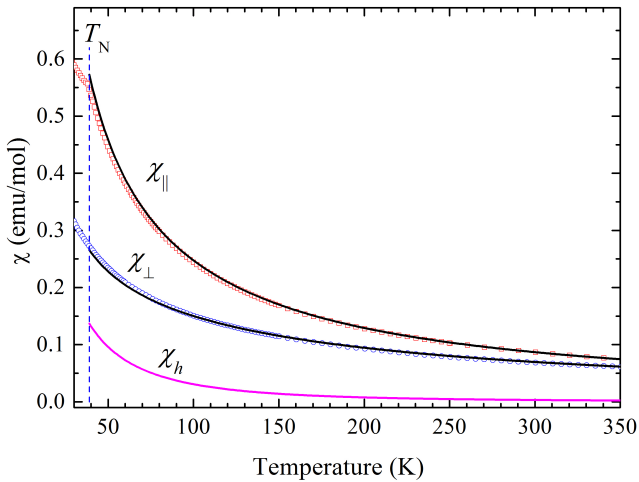

FIG. 5. Temperature dependencies of the longitudinal and transversal dc-susceptibilities of $DyFe_3(BO_3)_4$ measured in the external magnetic field $B$=0.05 T [19] (symbols) and the helix single-site non-diagonal susceptibility of the $Dy^{3+}$ ions. The results of calculations are represented by solid lines.

The helix chirality of the local susceptibility can be characterized by a non-diagonal component of the susceptibility of a single $Dy^{3+}$ ion, $\chi_h = N_A \chi_{Dy,yz}$. It determines a projection of the dysprosium magnetic moment on the $ab$-plane, induced by a magnetic field directed along the $c$-axis. This in-plane component of the dysprosium magnetic moment rotates around the magnetic field **B**//c in a clockwise (counterclockwise) direction in the $P3_121$ ($P3_221$) phase. The susceptibility $\chi_h$ is compared with $\chi_\parallel$ and $\chi_\perp$ in Fig. 5. The helix chirality of the local susceptibility could be detected in polarized neutron diffraction on paramagnetic $DyFe_3(BO_3)_4$ subjected to a magnetic field directed along the $c$ axis [45].

### C. Quadrupole moments of $Dy^{3+}$ in the $P3_121$ ($P3_221$) phase

The low-symmetry CF $C_2$ component distorts the electronic density distribution in the $Dy^{3+}$ ions and induces non-zero components of the quadrupole moment, in particular, $Q_{x_l^2-y_l^2} = \sum_j [C_2^{(2)}(j) + C_{-2}^{(2)}(j)]/2$ and $Q_{y_l z_l} = -i\sum_j [C_1^{(2)}(j) + C_{-1}^{(2)}(j)]/2$ (in units of $e<r^2>$) at the reference dysprosium site [46,47]. The quadrupole helix chirality appears due to rotations of the local $C_2$ symmetry axis by $2\pi/3$ and $4\pi/3$ at the $3a$ dysprosium sites shifted relative one another along the $c$-axis by $c/3$ and $2c/3$, respectively. The results of calculations of the average values of the quadrupole moment components,

$$<Q> = \mathrm{Tr}[Q\exp(-H_0/k_B T)]/\mathrm{Tr}[\exp(-H_0/k_B T)], \quad (12)$$

at different temperatures $T$ ($k_B$ is the Boltzman constant) using the final set of the CF parameters from column 5 in Table III in the single-ion Hamiltonian (1) agree satisfactorily with the experimental data presented in Ref. [3] (see Table IV) and confirm a supposition made in Ref. [3] that changes of $<Q_{x_l^2-y_l^2}>$ and $<Q_{y_l z_l}>$ with temperature are induced, mainly, by a redistribution of population between CF sublevels of the ground multiplet of the $Dy^{3+}$ ion.

### V. CONCLUSION

To summarize, we have performed a high-resolution polarized temperature-dependent optical spectroscopy and theoretical studies of $DyFe_3(BO_3)_4$ single crystals. Those are crystals, in which a new effect of "chirality of electronic quadrupole moments" was observed in resonant x-ray diffraction studies using circularly polarized x-rays [3]. We have shown that the low-symmetry crystal-field parameters for the $Dy^{3+}$ ion in the $P3_121$ ($P3_221$) phase, obtained from the crystal-field calculations done on the base of the spectral data, permit one to account quantitatively for this effect. The single-site magnetic susceptibility tensors of the magnetically nonequivalent $Dy^{3+}$ ions also have a property of helix chirality. As follows from calculations, transversal and longitudinal components of the dysprosium magnetic moment in an external magnetic field parallel to the $c$-axis have comparable values. The longitudinal magnetization is measured by a conventional magnetometric technique but the transversal magnetization, being a geometric sum of contributions connected by the ($2\pi/3$) rotations around the *c*-axis, equals zero. However, the rotating around the *c*-axis components of the single-ion magnetic moments in *ab*-planes can be found from the polarized neutron scattering data.

**ACKNOWLEDGMENTS**

This work was supported by the Russian Science Foundation (Grant №14-12-01033). I. A. G. thanks L. N. Bezmaternykh for useful comments concerning the crystal growth and the Russian Foundation for Basic Research for a financial support under Grant №14-02-00307a.